\documentclass[reprint,eqsecnum,floats,aps,amsmath,amssymb,nofootinbib,prd,onecolumn, showpacs]{revtex4-1}

\usepackage{graphicx}
\usepackage{amsmath,amssymb}
\usepackage{hyperref}
\usepackage{graphicx}
\usepackage{subfigure}
\usepackage{arydshln}
\usepackage{inputenc}
\usepackage{graphicx}
\usepackage{amsmath,amssymb}
\usepackage{hyperref}

\usepackage{mathtools}
\usepackage{float}

\DeclarePairedDelimiter{\abs}{\lvert}{\rvert}

\begin{document}

\title{Non-oscillating vacuum states and the quantum homogeneity and isotropy hypothesis in Loop Quantum Cosmology} 

\author{Beatriz Elizaga Navascu\'es}
\email{w.iac20060@kurenai.waseda.jp}
\affiliation{JSPS International Research Fellow, Department of Physics,
Waseda University, 3-4-1 Okubo, Shinjuku-ku, 169-8555 Tokyo, Japan}
\author{Guillermo  A. Mena Marug\'an}
\email{mena@iem.cfmac.csic.es}
\affiliation{Instituto de Estructura de la Materia, IEM-CSIC, Serrano 121, 28006 Madrid, Spain}
\author{Santiago Prado}
\email{santiago.prado@iem.cfmac.csic.es}
\affiliation{Instituto de Estructura de la Materia, IEM-CSIC, Serrano 121, 28006 Madrid, Spain}

\begin{abstract}
We study the compatibility of the quantum homogeneitiy and isotropy hypothesis (QHIH), proposed by Ashtekar and Gupt to restrict the choice of vacuum state for the cosmological perturbations in Loop Quantum Cosmology (LQC), with the requirement that the selected vacuum should lead to a power spectrum that does not oscillate. We inspect in close detail the procedure that these authors followed to construct a set of states satisfying the QHIH, and how a preferred vacuum can be determined within this set. We find a step that is not univocally specified in this procedure, in relation with the replacement of the set of states that was originally allowed by the QHIH with an alternative set that is more manageable. In fact, the first of these sets does not contain the state that has been used in most of the implementations of the QHIH to the analysis of the power spectrum of the perturbations in LQC. We focus our attention on the original set picked out by the QHIH and investigate whether some of its elements may display a non-oscillatory behavior. We show that, to the extent to which the techniques used in this paper apply, this possibility is feasible.  Thus, the two aforementioned criteria for the physical restriction of the vacuum state in LQC are compatible with each other and not exclusive.

\end{abstract}


\maketitle

\clearpage\thispagestyle{empty}

\section{Introduction}

In order to extract useful predictions from a physical theory, sometimes it does not suffice to determine the dynamical equations that rule the evolution of the system and analyze their properties. Choosing initial conditions can be just as important as these dynamical laws, especially for scenarios that cannot be reproduced in a controlled way in a laboratory (such as e.g. gravitational systems). In these cases, any successful theory should incorporate a procedure to determine suitable initial conditions, based on reasonable justifications and leading to phenomenologically sensible results. A situation in which this issue is particularly important is in the study of the evolution of primordial cosmological perturbations. These perturbations are believed to be the seeds of the temperature anisotropies that can be observed in the Cosmic Microwave Background (CMB) \cite{structures,mukhanov1,planck,planck-inf}. Beyond the standard cosmological paradigm, there is a growing hope that the power spectrum of the CMB radiation may keep some traces of the quantum geometry phenomena that would have taken place when the Universe was extremely young, and in this manner provide a way to falsify the predictions of quantum cosmology formalisms that describe the behavior of those early epochs \cite{Turok,AshtPRLLast,Ashtekarlast,Ivanlast}. 

Many attempts have been made to include quantum gravity effects in the analysis of primordial perturbations in cosmology (see e.g. \cite{HalliwellHawking,pintoneto1,pintoneto2, AshLewaDress,Bojo0,Bojo1,CLB,Bojo2,hyb-pert1,AAN3,Kiefer1,hyb-pert2,AAN1,AAN2,hyb-pert-eff,hyb-pert3,hyb-pert4,hyb-pert5,hybr-ten,Edward}). Most of these works adopt a Fock representation for the linear fields corresponding to the gauge-invariant perturbations. Then,  the choice of initial conditions is equivalent to selecting a preferred Fock vacuum state. A natural starting point is to demand that this vacuum state remains invariant under the spatial isometries of the homogeneous background (either treated as a classical or a quantum entity). Nonetheless, since the cosmological background is not stationary, these symmetries are usually not enough to pick out a unique vacuum, but only to restrict the choice within a family of states, unitarily equivalent among them in an optimal scenario, if the selection criteria have been wisely imposed \cite{uniquenessflat,uniquenessrep}. Within that family, one of the most common proposals is to choose the Bunch-Davies vacuum at the onset of inflation, especially if there is an inflationary phase that admits a slow-roll description. The Bunch-Davies state is arguably the most natural vacuum in de Sitter spacetime, which is believed to provide a good approximation for the cosmological expansion in such an inflationary phase \cite{mukhanov1,BD}. However, this state is not well adapted to the cosmological evolution if there are relevant regimes previous to slow-roll inflation with physical phenomena that can affect the primordial perturbations. For instance, this may happen for perturbation modes with wavelengths of the order of the characteristic scales associated with the quantum gravity processes that may have affected the Universe well before inflation, out of the domains of applicability of classical General Relativity. To take those primeval epochs into account, we must have some level of understanding of the underlying quantum geometry. Several candidate formalisms have been suggested to describe quantum gravity regimes in cosmology. Among them, we will focus our attention on Loop Quantum Cosmology (LQC)\cite{abl,ashparam,lqc1}, which is a nonperturbative quantization of cosmological systems based on the background-independent canonical theory of Loop Quantum Gravity \cite{LQG1,LQG2}. In LQC, for certain quantum states with interesting classical properties at large volumes, the Big Bang singularity becomes replaced with a quantum bounce \cite{APS,mmo}. 

Within these bouncing regimes of LQC, the choices of vacuum state for the perturbations that were first employed to extract predictions from the theory correspond to the so-called adiabatic states \cite{AAN1,AANv}. Adiabatic states \cite{adiabatic1,adiabatic2} are constructed iteratively from a zeroth-order state and, for sufficiently high order, they have the physically appealing property of permitting the renormalization of the stress-energy tensor. The adiabatic iterative process, however, is not mathematically robust and breaks down in certain circumstances. In addition, the motivation for using adiabatic conditions around a bounce of quantum origin is not completely clear from a theoretical point of view \cite{Wang1,Universe,Wang2}. A more recent proposal for the choice of a vacuum, with a more elaborated motivation, has been given by Ashtekar and Gupt \cite{AGvacio1,AGvacio2} in the context of the so-called dressed-metric approach to the study of primordial perturbations in LQC (see e.g. Refs. \cite{AAN3,AAN1,AAN2,Ivan}). According to this proposal, one chooses the state with a maximal classical behavior at the end of inflation among those that fulfill the so-called quantum homogeneity and isotropy hypothesis (QHIH). This is an extension into the quantum realm of Penrose's Weyl curvature hypotesis \cite{Penrose1,Penrose2}, which states that the Weyl curvature should vanish at the Big Bang.  The vacuum state selected so far using this QHIH has been seen to lead to a primordial power spectrum that is highly oscillatory in the dressed-metric approach to LQC, with respect to the wavenumber of the Fourier modes of the perturbations \cite{AGvacio2}. When these oscillations are suitably averaged, this power spectrum shows good agreement with the current CMB observations and it may even provide a way to alleviating certain anomalies reported by the Planck satellite \cite{AshtPRLLast,Ashtekarlast}. Nonetheless, it has been argued that these oscillations might come from an evolution of the proposed vacuum in the pre-inflationary epoch that blurs the information about the genuine effects of the LQC bounce on the perturbations \cite{no,NO-cond}. To deal with the problem of these superimposed oscillations in the power spectrum, Mart\'{\i}n de Blas and Olmedo put forward an alternative proposal, implemented numerically, which selects a vacuum state with a non-oscillatory (NO) spectrum \cite{no}. This NO-vacuum was originally introduced in the context of the so-called hybrid approach to LQC (see Ref. \cite{hlqc} for a comprehensive review on the topic). Recent investigations have identified some analytical conditions that must be satisfied by a vacuum displaying NO properties and that restrict its asymptotic behavior for infinitely large wavenumbers \cite{NO-cond,msdiag}.

The aim of this work is to investigate the relationship and compatibility between the QHIH and the NO-proposal as two criteria to restrict the choice of vacuum state in the context of hybrid LQC. In order to do this, we start by revisiting the mathematical conditions that define the admissible states according to the QHIH in a bouncing quantum cosmological scenario, indicating the steps where there appear ambiguities when the original proposal of Ashtekar and Gupt \cite{AGvacio1} is put into practice \cite{AGvacio2}. This construction starts by defining a ball of states that satisfies the QHIH in an interval around the quantum bounce, that is regarded as the Planck regime. In more detail, this interval is defined, for the sake of concreteness, as the period in which the density of the Universe is higher than $10^{-4}$ in Planck units.  Then, according to Ashtekar and Gupt, a preferred state should be selected within this ball such that it has maximal classical behavior at the end of inflation. However, because of the numerical complications in the imposition of these requirements, an alternative but much more manageable definition of the ball of admissible states was finally adopted in Ref. \cite{AGvacio2}. In the present work we show that this alternative definition leads to a different set of states to the original QHIH ball, raising the question of whether the Ashtekar-Gupt state selected in this manner actually lives in such original ball. We find that the answer is in the negative, at least in the hybrid approach to LQC. We recall that this state has highly oscillatory properties in the case of the dressed-metric approach to LQC, and it can be reasonably expected that it also displays this behavior in hybrid LQC\footnote{This is because both approaches share a classical pre-inflationary period that tends to produce oscillations in the evolution of the vacuum state, if this vacuum is not carefully chosen \cite{Universe}.}. Our result then opens the possibility that the preferred vacuum state that would arise from the original QHIH considerations may be compatible with the demand of an NO-behavior. With this motivation in mind and using the results of Ref. \cite{NO-cond}, we derive certain necessary conditions for the simultaneous satisfaction of an NO-behavior and the QHIH. Employing these conditions, we then examine the possibility that the Ashtekar-Gupt proposal may eventually lead to the choice of an NO-vacuum. Our result is that, without additional inputs, the two proposals are compatible. 

The structure of this paper is as follows. In Sec. II we take a close look at the definition of the QHIH given in Ref. \cite{AGvacio1} and the Ashtekar-Gupt vacuum selected in Ref. \cite{AGvacio2}. In Sec. III we identify a loose step in the passage from the theoretical construction of the ball of QHIH states, where this vacuum should reside, to its practical implementation. Furthermore, we derive certain compatibility conditions between the ball introduced in Ref. \cite{AGvacio1} and an NO-behavior. Section IV considers the analog in the hybrid approach to LQC of the Ashtekar-Gupt vacuum that was finally selected in Ref. \cite{AGvacio2}, proving that it does not belong to the original QHIH ball of states of Ref. \cite{AGvacio1}. In view of this result, in this section we also study the compatibility conditions between the original QHIH and an NO-behavior in hybrid LQC, showing that they are not exclusive. Section V contains our conclusions and further comments. Throughout this paper, we work in Planck units, setting $\hbar=c=G=1$.

\section{Construction of the Ashtekar-Gupt vacuum}

Given a real-valued function $s(\eta)$, where $\eta$ is a time coordinate, let us consider all complex solutions of the following family of differential equations: \begin{equation}\label{eq: ecms}
\mu_k^{\prime\prime}+(k^2+s)\mu_k=0,\quad k \in \mathbb{R}^{+} ,
\end{equation} 
that satisfy the normalization condition \begin{equation}\label{eq: condicionnormalizacion}
\mu_k\bar{\mu}_k^{\prime}-\mu_k^{\prime}\bar{\mu}_k=i.
\end{equation} Here, the prime denotes the derivative with respect to the time $\eta$ and the overhead bar indicates complex conjugation. On the other hand, let us consider a purely inhomogeneous real scalar field on $\mathbb{R}^4$ with Fourier coefficients labeled by a real wavevector $\vec{k}\in\mathbb{R}^3-\{\vec{0}\}$ and satisfying Eq. \eqref{eq: ecms} with $k=|\vec{k}|$. It is well known that any complete set $\{\mu_k\}_{k\in\mathbb{R}^{+}}$ of normalized solutions univocally defines a quantum Fock representation of the considered real scalar field \cite{Wald}.  Now, since Eq. \eqref{eq: ecms} is linear and real, we may write its general complex solution as a linear combination of a particular solution and its complex conjugate [which are functionally independent in virtue of Eq. \eqref{eq: condicionnormalizacion}]. It then follows that any two choices of basis elements, $\tilde\mu_k$ and $\mu_k$, may be related to each other through a linear Bogoliubov transformation:
\begin{equation}\label{eq: p0}
\tilde\mu_k\left(\eta\right)=\alpha_k\left(\tilde\mu_k,\mu_k\right)\mu_k\left(\eta\right)+\beta_k\left(\tilde\mu_k,\mu_k\right)\bar{\mu}_k\left(\eta\right).
\end{equation} 
The normalization condition \eqref{eq: condicionnormalizacion} holds provided that the constant Bogoliubov coefficients satisfy \begin{equation}\label{eq: condicionalfabeta}
\vert\alpha_k\left(\tilde\mu_k,\mu_k\right)\vert^2-\vert\beta_k\left(\tilde\mu_k,\mu_k\right)\vert^2=1,\quad  \forall k\in \mathbb{R}^+ .
\end{equation}
A choice of solutions $\{\mu_k\}_{k\in\mathbb{R}^{+}}$ is often called a basis of positive-frequency solutions, and it completely specifies a vacuum state from which the Fock space can be constructed. Henceforth, we refer to the vacuum state selected by a specific basis $\{\mu_k\}_{k\in\mathbb{R}^{+}}$ as $|0_{\mu}\rangle$.

In cosmological perturbation theory, both classically as well as for several approaches to quantum cosmology, an equation of the form \eqref{eq: ecms} typically dictates the propagation of the mode coefficients of the real Mukhanov-Sasaki field that describes the gauge-invariant scalar perturbations \cite{MukhanovSasaki,sasaki,sasakikodama}. Therefore, this equation is frequently called the Mukhanov-Sasaki equation.  Moreover, the dynamics of the tensor perturbations is also ruled by an equation of this type \cite{mukhanov1}. The function $s(\eta)$ is commonly referred to as the (effective) mass of the perturbations, and it can be given as a function of the geometrical variables of the background on classical solutions, or alternatively on quantum background states \cite{AAN1,hyb-pert4}. In this context, any choice of basis of positive-frequency solutions (or, equivalently, of their initial conditions) amounts to the choice of a specific vacuum state in the Fock quantization of the perturbations. 

Let us define $\mu_k^{\eta_0}$ as the solution to Eq. \eqref{eq: ecms} determined by the following initial conditions at any given time $\eta_0$:
\begin{equation}
\mu_k^{\eta_0}(\eta_0)=\frac{1}{\sqrt{2k}},\quad \mu_k^{\eta_0\prime}(\eta_0)=-i\sqrt{\frac{k}{2}}.
\end{equation} 
The basis constructed from such solutions $\mu_k^{\eta_0}$ for all $k$ gives rise to the so-called adiabatic state of zeroth-order, $|0_{\mu^{\eta_0}}\rangle$ \cite{adiabatic2}. The family of adiabatic states of zeroth-order parameterized by $\eta_0$ has some interesting physical properties, as we have succinctly commented in the Introduction. One of these properties is that $|0_{\mu^{\eta_0}}\rangle$ is the unique state that exactly fulfills at time $\eta_0$ the QHIH formulated in Refs. \cite{AGvacio1,AGvacio2}. In the case of tensor perturbations, this condition can be understood as an instantaneous quantum generalization of Penrose's Weyl curvature hypothesis, that takes into account and minimizes the quantum uncertainties of the operators representing the Weyl tensor at $\eta_0$. Owing to the similarities between the dynamics of the tensor and scalar perturbations, the QHIH has also been proposed in order to select a preferred family of quantum states for the Mukhanov-Sasaki field \cite{AGvacio2}.

Actually, the space of states allowed in the analysis of Ashtekar and Gupt is larger than the family of adiabatic states of zeroth-order that we have introduced. This is to cope with the fact that the dynamical evolution of the quantum field states is non-trivial in cosmological scenarios, something that leads to an instability of the instantaneous QHIH condition as time evolves. Explicitly, the QHIH is fulfilled at time $\eta_0$ by a normalized solution $\tilde{\mu}_k$ of Eq. \eqref{eq: ecms} if and only if \cite{AGvacio1}
\begin{equation}\label{eq: lambdazeta}
\Lambda_k(\tilde{\mu},\eta_0)=1,\qquad	\Lambda_k(\tilde{\mu}_k,\eta)=k\abs*{ \tilde{\mu}_k(\eta)}^2+\frac{1}{k}\abs*{ \tilde{\mu}_k^{\prime}(\eta)}^2.
\end{equation} 
As commented above, this condition alone fixes $\tilde{\mu}_k=\mu_k^{\eta_0}$ (up to a constant phase that does not affect the definition of the corresponding vacuum). In fact, one may write $\Lambda_k$ in terms of beta coefficients for Bogoliubov transformations to adiabatic states,
\begin{equation}\label{eq: lambdabeta}
\Lambda_k(\tilde{\mu}_k,\eta)=1+2\vert\beta(\tilde{\mu}_k,\mu_k^\eta) \vert^2. 
\end{equation} 
This implies that $\Lambda_k(\mu_k^{\eta_0},\eta)> 1$ in general, for any $\eta\neq \eta_0$. In view of this property of the cosmological system, Ashtekar and Gupt generalized the instantaneous QHIH condition to a dynamical one by requiring that physically admissible vacuum states should belong to the set \cite{AGvacio1,AGvacio2}
 \begin{equation}\label{eq: bolatotaldef}
B=\left\{|0_{\tilde{\mu}}\rangle\ \bigg\vert     \Lambda_k(\tilde{\mu}_k,\eta)\leq z_k,\ \forall k\in\mathbb{R}^+,\eta\in I\right\},
\end{equation} 
where $I$ is certain compact interval of time and we have defined the supremum\footnote{This definition is consistent as long as $s(\eta)$ has no singularities in $I$. This is the case for LQC, where $s(\eta)$ is obtained from well-defined expectation values of quantum geometry operators.}
\begin{equation}\label{eq: thesupremes}
z_k=\sup_{\eta_0,\eta_1\in I}\Lambda_k(\mu_k^{\eta_0},\eta_1).
\end{equation} 
We will refer to the family of states $B$ as the \textit{total} Weyl uncertainty ball.  Obviously, its construction depends on the choice of interval $I$. Since the QHIH is a generalization of Penrose's Weyl curvature hypothesis, which should only be applied in the high-curvature regime of spacetime, it is natural to demand that this interval coincides with the period where important quantum cosmological phenomena take place (the so-called Planck regime). Specifically, in Ref. \cite{AGvacio1} this interval was defined as the epoch in which the density of the Universe is higher than $10^{-4}$ Planck units.

In order to extract robust physical predictions from the theory, one needs to single out a preferred vacuum state within $B$ by demanding a suitable behavior. According to Ref. \cite{AGvacio2}, this preferred state must minimize the quantum dispersions of the field operators at the end of inflation, so that the state has optimal classical properties at times when the quantum effects should be negligible. In practice, it is a complicated task to find such a state starting from $B$ (even from a numerical perspective). This difficulty was circumvented in Ref. \cite{AGvacio2} by instead searching for the vacuum among states that live in \textit{instantaneous} Weyl uncertainty balls $B_{\eta_0}$, defined as follows:
\begin{equation}
B_{\eta_0}=\left\{|0_{\tilde{\mu}}\rangle \ \bigg\vert\ \Lambda_k(\tilde{\mu}_k,\eta_0)\leq z^{\eta_0}_k \,\ \forall k\in\mathbb{R}^+\right\},
\end{equation} 
where
\begin{equation}
z^{\eta_0}_k = \sup_{\eta\in I}\Lambda_k(\mu_k^{\eta},\eta_0).
\end{equation}
According to Ref. \cite{AGvacio2}, the state $|0_{\nu^{\eta_0}}\rangle$ that minimizes the quantum field dispersions at the end of inflation, within the instantaneous Weyl uncertainty ball $B_{\eta_0}$, has the form\footnote{We are greatly thankful to A. Wang, P. Singh, B. Li, and T. Zhu for pointing out a previous misprint in this formula to us.}
\begin{equation}\label{nueq}
\nu_k^{\eta_0}(\eta)=\sqrt{1+\left(r_k^{\eta_0}\right)^2}\mu_k^{\eta_0}(\eta)+r_k^{\eta_0}e^{-i\theta_k^{\eta_0}}\bar{\mu}{}_k^{\eta_0}(\eta),
\end{equation} 
where 
\begin{equation}\label{classcond}
\left( r_k^{\eta_0}\right)^2=\frac{1}{2}\left( z_k^{\eta_0}-1\right), \quad\quad \theta_k^{\eta_0}=\pi-2\arg\left[\mu_k^{\eta_0}(\eta_{\text{end}})\right],
\end{equation} 
where $\eta_{\text{end}}$ marks the end of inflation, $r_k^{\eta_0}\geq 0$, and $\arg$ denotes the argument of the complex quantity. Considering then all instantaneous Weyl uncertainty balls in the Planck regime, we have a one-parameter family of states that minimize the  quantum dispersions at $\eta_{\text{end}}$. The state corresponding to the global minimum is the unique Ashtekar-Gupt vacuum\footnote{In principle, there is no guarantee that there exists such global minimum simultaneously for all $k$. If this did not happen, one may instead choose the state $|0_{\nu^{\eta_0}}\rangle$ that minimizes a (suitably defined) average of the quantum dispersions over all of the modes. Alternatively, one may construct a new state by picking out each positive-frequency solution, among the two-parameter family $\{\nu_k^{\eta_0}\}$, that minimizes the quantum dispersions for each $k$ separately. Unfortunately, by its construction, one cannot generally assure that the state that would result from this last procedure belongs to any of the instantaneous balls $B_{\eta_0}$. So, we will not consider this possibility in this paper.}.

\section{QHIH: Ambiguities in its implementation and combination with the NO-proposal}
\subsection{Difference between balls of states}

As we have commented, the actual construction of the Ashtekar-Gupt vacuum state put forward in Ref. \cite{AGvacio2} does not start from the total Weyl uncertainty ball of states $ B$, but rather from the union of instantaneous balls, $ \bigcup_{\eta_0 \in I} B_{\eta_0}$.  An important question that immediately arises is whether the two sets of states are equal. If this were the case, then the procedure followed in Ref. \cite{AGvacio2} to find the state with a maximally classical behavior at the end of inflation would be, without question, consistent with the QHIH originally proposed in Ref. \cite{AGvacio1} (and actually used as a motivation in Ref. \cite{AGvacio2}). In the following, we show that the answer is in the negative.

We begin by using Eq. \eqref{eq: lambdazeta} to rewrite the definition of $ B$ and $ B_{\eta_0}$ in terms of beta coefficients,
\begin{align}\label{eq: bolabetaeta}
B_{\eta_0}&=\left\{|0_{\tilde{\mu}}\rangle\ \Bigg\vert\ \abs*{\beta_k(\tilde{\mu}_k,\mu_k^{\eta_0})}^2\leq  \sup_{\eta\in I}\abs*{\beta_k(\mu_k^{\eta},\mu_k^{\eta_0})}^2\quad \forall k \in \mathbb{R}^+ \right\},\\
B&=\left\{|0_{\tilde{\mu}}\rangle\ \Bigg\vert\ \abs*{\beta_k(\tilde{\mu}_k,\mu_k^\eta)}^2\leq  \sup_{\eta_0,\eta_1\in I}\abs*{\beta_k(\mu_k^{\eta_0},\mu_k^{\eta_1})}^2\quad \forall k \in \mathbb{R}^+,\forall \eta\in I \right\}.
\end{align} 
Given a compact interval $I$ and any positive $k$ there exist times $\eta_{-}^{k}$ and $\eta_{+}^{k}$ in $I$ such that 
\begin{equation}
\abs*{\beta_k\left(\mu_k^{\eta_{-}^{k}},\mu_k^{\eta_{+}^{k}}\right)}^2=\sup_{\eta_0,\eta_1\in I}\abs*{\beta_{k}\left(\mu_k^{\eta_0},\mu_k^{\eta_1}\right)}^2.
\end{equation}
Considering a \emph{fixed} (but otherwise generic) wavenumber $\tilde{k}$, let us then define a state $|0_{\mu^S}\rangle$ such that
\begin{align}
\mu_k^S\left(\eta\right)=\bar{\alpha}_k\left(\mu_k^{\eta_{+}^{\tilde{k}}},\mu_k^{\eta_{-}^{\tilde{k}}}\right)\mu_k^{\eta_{+}^{\tilde{k}}}\left(\eta\right)+\beta_k\left(\mu_k^{\eta_{+}^{\tilde{k}}},\mu_k^{\eta_{-}^{\tilde{k}}}\right)\bar{\mu}_k^{\eta_{+}^{\tilde{k}}}\left(\eta\right). \label{eq: p1}
\end{align} 
This state belongs to the instantaneous ball $B_{\eta_{+}^{\tilde{k}}}$, which is by definition contained in the union $\bigcup_{\eta_0 \in I} B_{\eta_0}$. In order to show this, we first notice that
\begin{equation}
\abs*{\beta_{k}\left(\mu_{k}^S,\mu_{k}^{\eta_{+}^{\tilde{k}}}\right)}=\abs*{\beta_{k}\left(\mu_{k}^{\eta_{-}^{\tilde{k}}},\mu_{k}^{\eta_{+}^{\tilde{k}}}\right)},
\end{equation}
as one can check using the general property $\vert\beta_k\left(\tilde\mu_k,\mu_k\right)\vert=\vert\beta_k\left(\mu_k,\tilde\mu_k\right)\vert$, which follows from Eqs. \eqref{eq: p0} and \eqref{eq: condicionalfabeta}. Thus, for any $k$ and taking into account the definition of supremum, it holds that
\begin{equation}
\abs*{\beta_{k}\left(\mu_{k}^S,\mu_{k}^{\eta_{+}^{\tilde{k}}}\right)}^2\leq \sup_{\eta\in I} \abs*{\beta_{k}\left(\mu_{k}^{\eta},\mu_{k}^{\eta_{+}^{\tilde{k}}}\right)}^2.
\end{equation}
This inequality can at most be saturated, as it happens e.g. for $k=\tilde{k}$. Hence we conclude that, according to the definition of instantaneous ball given in Eq. \eqref{eq: bolabetaeta}, the state $|0_{\mu^S}\rangle$ belongs to $B_{\eta_{+}^{\tilde{k}}}$ as we wanted to show.

Now, we can write the basis element $\mu^{\eta_{+}^{\tilde{k}}}_k$ in terms of the Bogoliubov coefficients that relate it to $\mu^{\eta_{-}^{\tilde{k}}}_k$,
\begin{equation}\label{eq: p2}
\mu_k^{\eta_{+}^{\tilde{k}}}\left(\eta\right)=\alpha_k\left(\mu_k^{\eta_{+}^{\tilde{k}}},\mu_k^{\eta_{-}^{\tilde{k}}}\right)\mu_k^{\eta_{-}^{\tilde{k}}}\left(\eta\right)+\beta_k\left(\mu_k^{\eta_{+}^{\tilde{k}}},\mu_k^{\eta_{-}^{\tilde{k}}}\right)\bar{\mu}_k^{\eta_{-}^{\tilde{k}}}\left(\eta\right).
\end{equation} 
Composing the transformations \eqref{eq: p2} and  \eqref{eq: p1}, we see that
\begin{align}
\beta_k\left(\mu_k^S,\mu_k^{\eta_{-}^{\tilde{k}}}\right)=2\bar{\alpha}_k\left(\mu_k^{\eta_{+}^{\tilde{k}}},\mu_k^{\eta_{-}^{\tilde{k}}}\right)\beta_k\left(\mu_k^{\eta_{+}^{\tilde{k}}},\mu_k^{\eta_{-}^{\tilde{k}}}\right).
\end{align} 
Therefore, focusing our discussion on the mode $k=\tilde{k}$, we have that
\begin{eqnarray}
\abs*{\beta_{\tilde{k}}\left(\mu_{\tilde{k}}^S,\mu_{\tilde{k}}^{\eta_{-}^{\tilde{k}}}\right)}^2=4\abs*{\alpha_{\tilde{k}}\left(\mu_{\tilde{k}}^{\eta_{+}^{\tilde{k}}},\mu_{\tilde{k}}^{\eta_{-}^{\tilde{k}}}\right)}^2\abs*{\beta_{\tilde{k}}\left(\mu_{\tilde{k}}^{\eta_{+}^{\tilde{k}}},\mu_{\tilde{k}}^{\eta_{-}^{\tilde{k}}}\right)}^2 \geq 4 \sup_{\eta_0,\eta_1\in I}\abs*{\beta_{\tilde{k}}\left(\mu_{\tilde{k}}^{\eta_0},\mu_{\tilde{k}}^{\eta_1}\right)}^2 , \label{eq: BnoB}
\end{eqnarray} 
where we have used that the squared norm of the alpha-coefficient is never smaller than the unit because of the normalization condition \eqref{eq: condicionalfabeta}. This inequality straightforwardly implies that $|0_{\mu^S}\rangle$ does not belong to $ B$, and hence we have that $ B \neq \bigcup_{\eta_0 \in I} B_{\eta_0}$. Of course, this does not mean that the intersection of these two sets is empty. In fact, we clearly have that any adiabatic state of zeroth-order $|0_{\mu^{\eta}}\rangle$, with $\eta\in I$, automatically belongs to both sets.

\subsection{Non-oscillatory requirements for states in the Weyl uncertainty ball}

The primordial power spectrum of the perturbations in a state $|0_{\tilde{\mu}}\rangle$ can be obtained from the evaluation of $\abs*{\tilde{\mu}_k}^2$ at the end of slow-roll inflation. The dynamical evolution of the perturbations from their initial conditions in the Planck regime to this stage when inflation ends can leave imprints that are potentially observable in the CMB. In particular, any oscillatory behavior of the amplitude of the positive-frequency solutions during the pre-inflationary evolution may affect the spectrum and, in this way, produce oscillations in it. These oscillations may be superimposed to the genuine imprints of the pre-inflationary dynamics of the Universe on the spectrum, including quantum gravity modifications, and blur them \cite{NO-cond}. With this motivation in mind, Mart\'{\i}n de Blas and Olmedo proposed a criterion to select a state with non-oscillatory behavior, called the NO-vacuum, which minimizes the oscillations in the spectrum over the interval between the time where the initial conditions are imposed and the onset of inflation \cite{no}. The implementation of this criterion was generally numerical, in the way in which it was originally introduced. 

More recently, it has been possible to derive some necessary conditions that an NO-vacuum has to satisfy. In detail, given an NO-vacuum $|0_{\mu^{NO}}\rangle$ (the existence of which is supported at least from a numerical perspective), we can write the squared amplitude of the basis of positive-frequency solutions associated with any other state $|0_{\tilde{\mu}}\rangle$ as \cite{NO-cond},
\begin{equation}
\abs*{\tilde{\mu}_k}^2=\frac{1}{2}\abs*{\mu^{NO}_k}^2\left[A+B+ (A-B)\cos(2\phi_k)+2C\sin(2\phi_k)\right], \qquad \phi^{\prime}_k=\frac{1}{2}\abs*{\mu^{NO}_k}^{-2},
\end{equation}
where $A$, $B$, and $C$ are real constants, with $C^2=AB-1$. 
As long as there exists a sufficiently long regime in the evolution of the perturbations in which  $2\abs*{\abs*{\mu^{NO}_k}^{\prime} \abs*{\mu^{NO}_k}}<1$, it follows from this formula that any other NO-vacuum state must have constants $A$ and $B$ lying in a close neighbourhood of the unit. The existence of such a regime is expected in any pre-inflationary cosmological evolution that resembles the Einsteinian one for a universe with a massless scalar field at low energy densities, as it is the case e.g. in interesting LQC scenarios \cite{Ivan,Universe}. This is because, in low-curvature regimes of General Relativity where the energy density of the inflaton is dominated by its kinetic contribution, the mass $s(\eta)$ is a very slowly varying function of time \cite{hybr-pred,NO-cond}.

Following these considerations, we can regard as a necessary condition for any candidate to be an NO-vacuum $|0_{\mu^{NO}}\rangle$ that it must satisfy $2\abs*{\abs*{\mu^{NO}_k}^{\prime} \abs*{\mu^{NO}_k}}<1$ at least for all times $\eta$ near the end of the Planck regime. Explicitly, if we write the basis of positive-frequency solutions $\{\mu_k^{NO}\}$ in the form \cite{NO-cond}
\begin{equation}\label{eq: cond1}
\mu_k^{NO}=\frac{1}{\sqrt{-2\text{Im}(h_k)}}e^{i\int_{\eta_{1}}^{\eta} d\tilde\eta\, \text{Im}(h_k)(\tilde\eta)},
\end{equation} 
where $\eta_1$ is a reference time, irrelevant for the choice of vacuum state, and $h_k$ is a solution to the Riccati equation 
\begin{equation}\label{eq: cond2}
h_k^{\prime}=k^2+s+h_k^2
\end{equation} 
with strictly negative imaginary part, then the aforementioned necessary condition on an NO-vacuum can be equivalently expressed as
\begin{eqnarray}\label{eq: cond3}
&&\abs*{\text{Re}(h_k)(\eta)}=\epsilon_k (\eta)\abs*{\text{Im}(h_k)(\eta)}\quad {\rm with} \quad 0<\epsilon_k (\eta)< 1, \\
\label{eq: cond4}
&&\left |\frac{k^2+s(\eta)}{\text{Im}(h_k)(\eta)}-[1+\epsilon_k^2(\eta)]\text{Im}(h_k)(\eta) \right |< 1,
\end{eqnarray} 
at least for all times $\eta$ at the end of the Planck regime. Actually, we expect the above expressions involving $h_k$ to be much smaller than the unity, in particular for the resulting value of  $\epsilon_k$.

One can use the Bogoliubov transformation between the NO-vacuum and a zeroth-order adiabatic state $|0_{\mu^{\eta}}\rangle$ to obtain that
\begin{eqnarray}
\abs*{\mu^{NO}_k\left(\eta\right)}&=&\frac{1}{\sqrt{2k}}\abs*{\alpha_k\left(\mu^{NO}_k,\mu^{\eta}_k\right)+\beta_k\left(\mu^{NO}_k,\mu^{\eta}_k\right)},\nonumber\\ 
\abs*{\mu^{NO}_k{}^{\prime}\left(\eta\right)}&=&\sqrt{\frac{k}{2}}\ \abs*{\alpha_k\left(\mu^{NO}_k,\mu^{\eta}_k\right)-\beta_k\left(\mu^{NO}_k,\mu^{\eta}_k\right)}.
\end{eqnarray} 
These identities, combined with Eqs. \eqref{eq: cond2} and \eqref{eq: cond3}, imply that 
\begin{equation}\label{eq: agnocond0}
2\Lambda_k(\mu_k^{NO},\eta)=\frac{k}{\abs*{\text{Im}(h_k)(\eta)}}+[1+\epsilon_k^2(\eta)]\frac{\abs*{\text{Im}(h_k)(\eta)}}{k}
\end{equation} 
for times $\eta$ at the end of the Planck regime.

Hence, given an interval $I$ defining this Planck regime, a state satisfying the first necessary NO-vacuum condition \eqref{eq: cond3} can belong to the total Weyl uncertainty ball $ B$ only if
\begin{equation}
\frac{k}{\abs*{\text{Im}(h_k)(\eta)}}+[1+\epsilon_k^2(\eta)]\frac{\abs*{\text{Im}(h_k)(\eta)}}{k}\leq 2 z_k,
\end{equation} 
for all $\eta$ at the end of $I$ [and with $\epsilon_k(\eta)$ being small]. This inequality can be solved and leads to the following restriction:
\begin{equation}
[1+\epsilon_k^2(\eta)]\abs*{\text{Im}(h_k)(\eta)}\in \left[k z_k-k\sqrt{ z_k^2-[1+\epsilon_k^2(\eta)]},k z_k+k\sqrt{ z_k^2-[1+\epsilon_k^2(\eta)]}\right].
\end{equation}
On the other hand, the second necessary condition \eqref{eq: cond4} for an NO-vacuum is satisfied if and only if
\begin{equation}
[1+\epsilon_k^2(\eta)]\abs*{\text{Im}(h_k)(\eta)}\in \left(\frac{1}{2}\sqrt{1+4[k^2+s(\eta)][1+\epsilon_k^2(\eta)]}-\frac{1}{2},\frac{1}{2}\sqrt{1+4[k^2+s(\eta)][1+\epsilon_k^2(\eta)]}+\frac{1}{2}\right).
\end{equation}
Recalling that $\epsilon_k$ is expected to be much smaller than the unity for an NO-vacuum, at leading order we can ignore the contribution of this parameter in the above expressions. With this approximation, it follows that the two necessary conditions for an NO-vacuum can only be compatible with the QHIH (as formulated in terms of the total ball $ B$) if
\begin{equation}\label{eq: consistency}
\left[k z_k-k\sqrt{ z_k^2-1},k z_k+k\sqrt{ z_k^2-1}\right]\bigcap \left(\frac{1}{2}\sqrt{1+4[k^2+s(\eta)]}-\frac{1}{2},\frac{1}{2}\sqrt{1+4[k^2+s(\eta)]}+\frac{1}{2}\right)\neq \emptyset, 
\end{equation}
for all instants of time $\eta$ near the end of $I$. It is worth remarking that this interval $I$ should cover all of the Planck regime, so that it smoothly connects with a kinetically dominated universe where, according to General Relativity, the mass $s(\eta)$ varies very slowly over time.

One can similarly obtain a consistency requirement for a state that satisfies the necessary NO-vacuum conditions in order that it also belongs to the instantaneous Weyl uncertainty balls $B_{\eta_0}$ for times $\eta_0$ close to the end of $I$: it suffices to replace $\eta$ with $\eta_0$ and $ z_k$ with $ z_k^{\eta_0}$ in Eq. \eqref{eq: consistency}. Nonetheless, we will mainly focus our attention on the consistency of the non-oscillatory behavior with the QHIH formulated in terms of the total ball $B$. This is so because of two reasons. The first one is that this is the original formulation of the QHIH, motivated in Ref. \cite{AGvacio1} on fundamental issues. The second one is that, for the alternative formulation of the QHIH given in Ref. \cite{AGvacio2}, our current analytic knowledge of the NO-vacua only allows us to study the consistency requirement on solid grounds for instantaneous Weyl uncertainty balls $B_{\eta_0}$ defined at times $\eta_0$ that are near the end of the Planck regime.

\section{QHIH in hybrid LQC: Compatibility with the NO-proposal}\label{sec: hlqc}

Our previous discussion is valid for any choice of mass $s(\eta)$ for the perturbations, provided that it it is non-singular in the time interval of interest, and that it varies slowly at the end of this interval. In the following we will focus our attention on the case that this mass is given by the evaluation on effective LQC backgrounds of the result of a hybrid quantization of the perturbed inflationary cosmology. In this hybrid approach, the Friedmann-Lema\^{\i}tre-Robertson-Walker (FLRW) background is quantized according to LQC, while the perturbations are treated with typical techniques of quantum field theory in curved spacetimes, more specifically by adopting a Fock description. If we consider certain quantum states for the background in LQC that are highly peaked in bouncing trajectories, the evaluation of background operators on these quantum states can be well approximated by considering the evaluation of their classical analogs on the peak trajectories. Actually, these peak trajectories follow the evolution dictated by an effective Hamiltonian constraint on the FLRW background. In this background, inflation is driven by a homogeneous scalar field subject to a potential, that we will particularize to a quadratic one for simplicity. In this setting, any background solution is completely fixed by the value of the inflaton field at an arbitrary initial time, e.g. at the bounce, and of its mass $m$. From a phenomenological point of view, in order to obtain power spectra that are compatible with the observations but still are capable of including traces of the LQC effects, the typical effective solutions that turn out to be interesting present a classical era shortly after the bounce in which the kinetic energy of the inflaton greatly dominates over its potential, era that extends almost until the onset of inflation \cite{Ivan,Universe}. This type of solutions is obtained for initial values of the inflaton at the bounce and values of its mass close to  $\phi_B=0.97$ and $m=1.2\cdot 10^{-6}$, data that we will adopt from now on for our analyses \cite{hybr-pred,Universe}. In the hybrid approach, the gauge-invariant perturbations that propagate on the above LQC backgrounds follow dynamical equations of the form \eqref{eq: ecms}.  

We can numerically integrate  the background evolution with the aforementioned initial conditions to obtain the value of the mass $s(\eta)$, and then apply the procedure that we have explained in Sec. II to determine the Ashtekar-Gupt vacuum in hybrid LQC. For this numerical integration, we use Verner's ``Most Efficient" 9/8 Runge-Kutta method (with a lazy 9th-order interpolant) \cite{vern9,diffeq}. To implement this procedure, we first need to characterize the Planck regime in a precise manner. According to the definition given by Ashtekar and Gupt, which requires values of the inflaton energy density above $10^{-4}$, the conformal times that define the considered regime are $I_{PL}=[-4.2,4.2]$ (with $\eta=0$ corresponding to the bounce).  Employing the interval $I= I_{PL}$, we can obtain values of the upper bounds $ z_k^{\eta_0}$, for all $\eta_0\in I_{PL}$, and $ z_k$, which respectively define the instantaneous and total Weyl uncertainty balls $ B_{\eta_0}$ and $ B$. As in the case of the dressed-metric approach to LQC \cite{AGvacio1,AGvacio2}, these bounds rapidly approach the unit for Fourier scales $k$ that are much larger than the Planck scale, which is the characteristic order of magnitude of the spacetime curvature around the bounce in LQC. This behavior reflects the fact that the effects of the cosmological evolution on a zeroth-order adiabatic state are negligible in the ultraviolet regime, and hence the ultraviolet scales approximately remain in this vacuum state, thus satisfying the QHIH at all times. On the other hand, for scales of the Planck order and smaller, the effects of the cosmological evolution on the dynamics of a zeroth-order adiabatic state become increasingly important, and as a consequence the bounds $ z_k^{\eta_0}$ and $ z_k$ grow above one in the infrared regime.

\begin{figure}
	\centering
	\includegraphics[width=10cm]{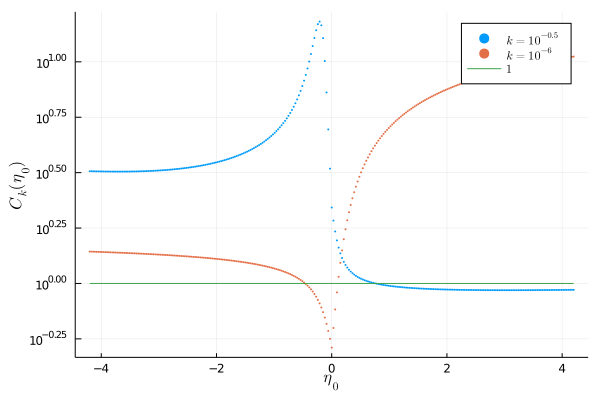}
	\caption{The quantity $C_k(\eta_0)=(z_k)^{-1} \max_{\eta\in I_{PL}}\Lambda_k(\nu_k^{\eta_0},\eta)$ compared with 1 for $k=10^{-6}$ and $k=10^{-0.5}$. There exists no value of $\eta_0$ such that the two curves remain below or equal to $1$, a fact that implies that no state $\nu_k^{\eta_0}$ lives in $B$.}
	\label{unbounded}
\end{figure}

With the obtained values of the bounds $ z_k^{\eta_0}$ that characterize the instantaneous uncertainty balls $ B_{\eta_0}$ in hybrid LQC, we can determine the initial conditions that correspond to the one-parameter family of states $|0_{\nu^{\eta_0}}\rangle$ with maximal classical behavior at the end of inflation. Indeed, in view of Eqs. \eqref{nueq} and \eqref{classcond} determining such states, the only additional data that we need are the phases of the adiabatic solutions $\mu_k^{\eta_0}$ at the end of inflation, which we compute numerically. As we explained in Sec. II, the vacuum state that Ashtekar and Gupt would propose as preferred, according to Ref. \cite{AGvacio2}, should lie in the resulting family of states. Recalling that the QHIH was originally formulated in terms of the total uncertainty ball $ B$ which, as we have shown in Sec. IIIA, is different to the union of instantaneous balls $\bigcup_{\eta_0\in I_{PL}}B_{\eta_0} $, the following question naturally arises: does the family of states $\{|0_{\nu^{\eta_0}}\rangle\}_{\eta_0 \in I_{PL}}$ actually belong to $B$, within the hybrid LQC framework? According to the definition given in Eq. \eqref{eq: bolatotaldef} for this total ball, the considered states belong to $ B$ if and only if, for each $\eta_0\in I_{PL}$, we have 
\begin{equation}
(z_k)^{-1} \max_{\eta\in I_{PL}}\Lambda_k(\nu_k^{\eta_0},\eta)\leq 1,
\end{equation}
for all $k\in \mathbb{R}^{+}$. In Fig. \ref{unbounded} we plot this function of $\eta_0$ for two representative values of $k$, showing that the above requirement cannot be met for any time $\eta_0$ in the Planck regime. Therefore, in the case of hybrid LQC, the vacuum state proposed by Ashtekar and Gupt in Ref. \cite{AGvacio2} does not belong to the total Weyl uncertainty ball that was originally motivated by the QHIH. From this perspective, the oscillatory behavior that is expected for this vacuum (taking into account its analog in dressed-metric LQC) does not imply an incompatibility between the NO-criterion and the original implementation of the QHIH when combined with a maximal classical behavior at the end of inflation.

\begin{figure}
\centering
\includegraphics[width=11cm]{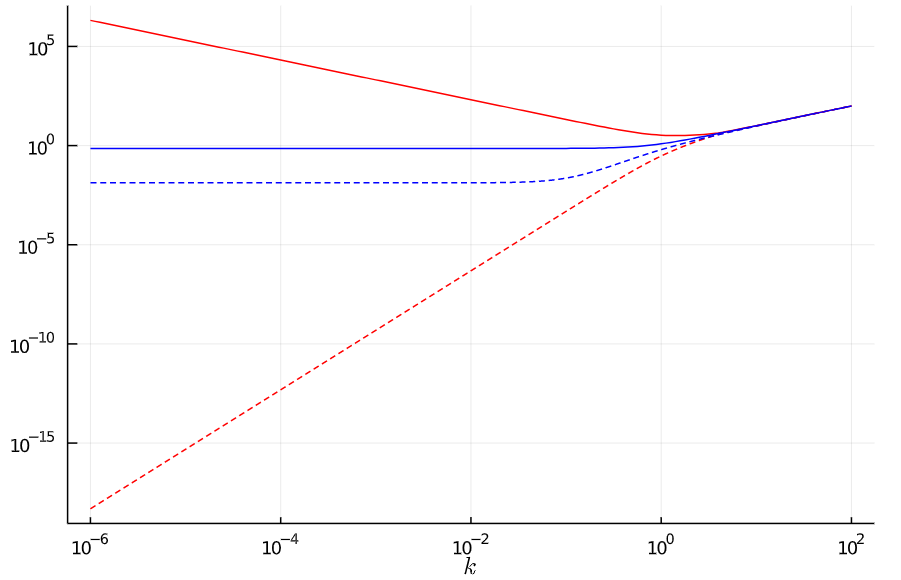}
\caption{The bounds imposed by the Weyl uncertainty ball, $ k \left(z_k-\sqrt{ z_k^2-1}\right)$ and $k \left( z_k+\sqrt{  z_k^2-1}\right)$, in red dashed and solid lines respectively, compared with the bounds imposed by the NO-condition,  $\frac{1}{2}\sqrt{1+4[k^2+s(\eta)]}-\frac{1}{2}$ and $\frac{1}{2}\sqrt{1+4[k^2+s(\eta)]}+\frac{1}{2}$, in blue dashed and solid lines respectively, for different modes $k$. These are evaluated for the mass $s(\eta)$ obtained in the hybrid approach to LQC, where $I=I_{PL}=[-4.2,4.2]$ and $\eta=4.2$, time near which the mass varies slowly. The intersection given by these bounds is not empty for any $k$.}
   \label{compatibility}
\end{figure}

This result further supports our decision to focus the attention on vacua belonging to the total uncertainty ball $ B$, when studying the compatibility of the QHIH with the necessary NO-vacuum conditions in hybrid LQC. According to our discussion in Sec. IIIB, in order to do this we just have to particularize the intervals appearing in Eq. \eqref{eq: consistency} to the case of hybrid LQC and check that their intersection is non-empty. All the ingredients needed for this test, namely the mass $s(\eta)$ and the upper bound $ z_k$, are readily available from our previous computations. In Fig. \ref{compatibility} we plot the curves that limit these two intervals at the representative time $\eta=4.2$ that marks the end of the Planck regime $I_{PL}$. We clearly see that their intersection is not empty, indicating the compatibility between the QHIH and the NO-criterion for the choice of a vacuum state of the cosmological perturbations in hybrid LQC. Actually, in the infrared regime (which is where oscillations can appear for the considered type of states \cite{Universe}) the $z_k$-independent interval related with the NO-condition is contained in the interval that corresponds to the original version of the QHIH. Therefore, it follows that any state that is an NO-vacuum satisfies this version of the QHIH at least at the end of the Planck regime.

\section{Conclusions}

With a combination of analytical and numerical means, we have investigated the compatibility between the QHIH proposed by Ashtekar and Gupt \cite{AGvacio1,AGvacio2} and the NO-proposal for the choice of initial conditions on primordial perturbations in quantum cosmology \cite{no,NO-cond}. In order to do this, we have examined in detail the construction that Ashtekar and Gupt employed to determine their vacuum state, and we have discussed a step that is not univocal and its consequences. In addition, we have derived some analytical conditions that a vacuum state must satisfy to comply with the original formulation of the QHIH introduced in Ref. \cite{AGvacio1} and with the NO-condition. We have followed the Asthekar-Gupt proposal, adapted to hybrid LQC, in the phenomenologically interesting case with a kinetically dominated pre-inflationary era. We have shown that the vacuum selected by the Ashtekar-Gupt construction in fact lies outside the ball of states that satisfy the QHIH according to the prescription of Ref. \cite{AGvacio1}, where it was motivated as a quantum generalization of Penrose's Weyl curvature hypothesis. Because of this, we have focused our attention on the properties of the states in this last ball, showing that the original formulation of the QHIH is perfectly compatible with an NO-behavior.

More specifically, our starting point has been a careful revisitation of the entire Ashtekar-Gupt proposal, applicable for any real-valued (and non-singular) function that plays the role of a time-dependent mass in the dynamical equations of the Fourier modes of the perturbations, which have the form of generalized harmonic oscillator equations. By translating Penrose's hypothesis to the quantum realm, one then defines a Weyl uncertainty ball for the states of the perturbations such that they all fulfill the QHIH in a specific interval of time \cite{AGvacio1}. Since this hypothesis is formulated for high-curvature regimes, this interval is chosen as the Planck regime in which quantum gravity effects are truly important. To extract meaningful predictions, a unique preferred state must be chosen from the ball obtained for the Planck interval: this is the state with a maximal classical behavior at times when quantum effects have become irrelevant, e.g. the end of inflation for concreteness. Finding this state directly with numerical methods is a very complicated task. As an alternative route, in Ref. \cite{AGvacio2} Ashtekar and Gupt opted to instead consider instantaneous Weyl uncertainty balls, defined at each instant of time in the studied interval. In this manner, one characterizes in an analytical way a one-parameter family of states, namely, one with maximal classical behavior for each instantaneous ball. Among them, one should numerically find the state with best classical properties and identify it with the Ashtekar-Gupt vacuum.

The need to replace the ball of states originally determined by the QHIH by its instantaneous counterparts can give rise to questions about the real habitat of the Ashtekar-Gupt vacuum, and to ambiguities in its construction, if these balls are different. In fact, in this work we have shown that the original Weyl uncertainty ball is actually different to the union of all its instantaneous counterparts. It is worth remarking that our proof is independent of the specific form of the mass of the perturbations, the choice of compact interval for the definition of the Planck regime, and the wavenumber of the Fourier mode. In addition, we notice that our proof does not exclude the fact that the two considered sets of states, even if different, have a non-empty intersection (e.g. adiabatic states of zeroth-order do belong to both sets). For a fixed functional form of the time-dependent mass of the perturbations, it is then legitimate to ask whether any of the states with maximal classical behavior in the instantaneous balls belongs to the original Weyl uncertainty ball. To get an answer in the case of the mass function derived in hybrid LQC, we have explicitly evaluated all these possible vacuum candidates of Ashtekar-Gupt type and shown that they do not belong to the ball obtained with the original implementation of the QHIH.

On the other hand, a physically relevant question one may ask to any viable choice of vacuum state is whether or not it leads to a highly oscillatory power spectrum. This oscillatory behavior can be considered an undesirable property inasmuch as it may blur away any modification to the primordial power spectrum that is ultimately caused by quantum geometry corrections \cite{NO-cond}. Actually, it is clear from the analysis carried out by Ashtekar and Gupt that these oscillations indeed appear in the dressed-metric approach for their choice of vacuum \cite{AGvacio2} (and a similar behavior can be expected for hybrid LQC). Our result shows, nonetheless, that in hybrid LQC this vacuum is outside of the ball of states that was introduced to comply with the fundamentals of the QHIH. This new perspective has led us to wonder whether the basic requirements on this set of admissible states are compatible with a non-oscillatory behavior of (at least) a subset of them. In order to answer this question, we have derived an analytical compatibility condition between these two types of requirements, and then have proceeded to check it in hybrid LQC. The result is satisfactory, from a theoretical perspective. Not only the QHIH is compatible with an NO-behavior but, furthermore, any NO-vacuum fulfills the QHIH
at least at the end of the Planck regime. 

The conclusions of this work are an important advance towards the theoretical motivation and determination of a preferred vacuum for the perturbations and the extraction of the corresponding physical predictions in (loop) quantum cosmology. Given two well-motivated criteria for the restriction of physically sound vacuum states (the QHIH proposed by Ashtekar and Gupt and the NO-criterion proposed by Mart\'{\i}n de Blas and Olmedo), both of which lead to predictions that are compatible with observations \cite{AshtPRLLast,no,hybr-pred}, testing and ensuring their compatibility is a key step to understand which physical conditions determine the quantum state of the primordial perturbations that explains the power spectrum that we observe nowadays in the CMB. This knowledge is paramount to investigate and falsify on a robust basis the phenomenological predictions that follow from any theory of quantum cosmology,  since an inappropriate choice of vacuum state can hide or misreflect the imprints that the genuine quantum cosmological dynamics may have left on the primordial fluctuations.

\acknowledgments

This work was partially supported by Projects No. MICINN FIS2017-86497-C2-2-P and No. MICINN PID2020-118159GB-C41 from Spain. B.E.N. acknowledges financial support from the Standard Program of JSPS Postdoctoral Fellowships for Research in Japan. S.P.  is grateful to Christopher Rackauckas and the Julia community for discussions regarding arbitrary numerical precision when defining differential equations.

\end{document}